\documentclass{article}
\usepackage{epsfig}
\begin{document}
\begin{center}
{{\bf \Large Research News --- Observation of Exotic Heavy Baryons 
}}

\bigskip

B. Ananthanarayan$^a$, 
Keshav Choudhary$^b$,
Lishibanya Mohapatra$^b$\\
Indrajeet Patil$^c$,
Avinash Rustagi$^b$,
K. Shivaraj$^a$

\bigskip
{\small
$^a$ Centre for High Energy Physics, Indian Institute of Science,
Bangalore 560 012\\
$^b$ St. Stephen's College, Delhi 110 007\\
$^c$ Fergusson College, Pune 411 004\\ 
}
\medskip
\end{center}


\medskip

\begin{abstract}
We review the recent discoveries of exotic heavy baryons at the
Fermi National Accelerator Laboratory.
\end{abstract}

\bigskip

According to a Press Release dated June 13, 2007 from the Fermi
National Accelerator Laboratory (Fermilab), the D0 Collaboration
has discovered a `triple-scoop' exotic heavy baryon, which has
been named $\Xi_b^-$(pronounced as `Zigh b') with mass of about 
$5.774\pm 0.019 {\rm GeV}/c^2$.  The Collaboration
has now posted its results on the internet, see~\cite{:2007ub}.  
Shortly afterwards, on June 15, the CDF Collaboration also at Fermilab
announced its discovery of the same particle, with a mass
$5.793\pm 0.003 {\rm GeV}/c^2$ (see Fermilab Press Release dated
June 25, 2007).  This is the first instance of production
of a baryon with three distinct
`flavours', {\it i.e.} with a quark from each of the three generations, and
represents a landmark in the history of elementary particle physics. 
The three generations consist of u(up) and d(down), c(charm) and s(strange), 
t(top) and b(bottom) quarks and each of the quarks carries an intrinsic angular
momentum or `spin' of $1/2$ in units of $\hbar$ ($\hbar = h/2 \pi$, where
$h$ is the Planck's universal constant). 
The $\Xi_b^-$ is made up of d, s and b quarks.
It should be noted here
that the first generation is made up of the lightest of 
the quarks,
and is the primary constitutent of all stable matter.
Somewhat earlier, another particle containing a single b- quark
and two light quarks, the $\Sigma_b$ was discovered
by the CDF Collaboration in Fermilab in 2006 ~\cite{hep-ex/0706.3868}. This 
highly unstable particle rapidly decays through the strong interactions with the
emission of a pion and a $\Lambda_b$.  

These discoveries are triumph for 
Quantum Chromodynamics (QCD), the presently accepted theory of strong
interactions whose earliest form was announced by Murray Gell-Mann in
Bangalore some 50 years ago, for a historical discussion see ref~\cite{Johnson}.
Although
it was not very certain what quarks meant at that time, we now know that they
are the microscopic degrees of freedom, along with gluons of the strong
interactions, which are confined within strongly interacting matter 
(hadrons) which come in two varieties {\it viz.} baryons and mesons. 
The conventional
protons and neutrons are examples of baryons while pions, which supply the
internucleon force, are examples of mesons. The former are made up of three
quarks, while the latter are made up of a quark and an anti-quark pair.
We now know that quarks come in six flavours, of which all but the u- quark
are unstable due to the presence of the weak interaction.  The latter
is responsible for the decay of a free neutron, and is the only force
that allows for the change in particle type.  At the time Gell-Mann 
proposed his hypothesis, it was necessary to consider only three types,
which are now called u, d and s quarks, 
carrying charges of $2e/3,\  -e/3$ and $-e/3$ respectively, where
e is the proton charge.  Both the s- and d- quarks can decay into 
a u- quark along with the emission of an electron and its anti-neutrino,
while the s- quark can also decay into a u- quark and a muon and
its anti-neutrino, where a muon is an elementary particle that is
like an electron except that it is about 210 times heavier.  The muon
and electron are charged leptons, recalling that
leptons are those particles that
do not `feel' the strong interactions, while neutrinos are what are called
neutral leptons.  Today we know that there is yet another 
more massive counterpart of the d- and s- quarks known as the b- quark.
Correspondingly, there are cousins of the u- quark known as c- and
t- quarks.  Correspondingly, therefore, the b- quark can decay into
both c- and u- quarks as it is heavier than each of them, and the
c-quark can decay into s- and d- quarks for the same reason.
This information will be of relevance in the latter part of this article.

Recalling that baryons are made up of quarks, one may readily see
that the lightest baryons, the proton and neutron are made up of uud and
udd quark combinations respectively. If one of the quarks in the proton or 
neutron is replaced by s quark in high energy physics experiments, where
s quark and anti-quark pairs are produced in collisions and the former 
is captured by, e.g., the
proton with the ejection of u or d quark, then one produces particles
which have been named $\Sigma$, e.g., $\Sigma^+$(uus), $\Sigma^0$(uds),
$\Sigma^-$(dds) while if the proton 
exchanges two of its intrinsic quarks for two s- quarks then one
produces particles which are named $\Xi$ , e.g., $\Xi^0$(uss), $\Xi^-$(dss).  
The results being reviewed
in this article concern those experiments in which, for the first time
baryons have been found with `s' as well as a `b' quark.
While existence of such particles may be considered a very natural state
of affairs in quark model of hadrons, their discovery actually proves to be a triumph of modern
experimental particle physics.  

Turning now to the specific discovery of the $\Xi_b^-$, these have been
produced in highly energetic collisions of protons and anti-protons
at the Fermilab Tevatron collider.  In the fireball ensuing in these
collisions there is a probability of producing many exotic states;
the probability of production of the states $\Xi_b$ in question is typically
small as the b- quark produced in such a fireball must be able to pick
up an s- quark in addition, and finally a light quark to form the baryon
of interest.  Then comes the question of identifying the production of
such particles through their decay chains such as shown in Fig 1. In this 
case, the b- quark
in the $\Xi_b^-$ decays through the production of a virtual $W^-$ which is
the messenger of weak force, and a c- quark.  
\begin{figure} [ht]
\centering
\includegraphics{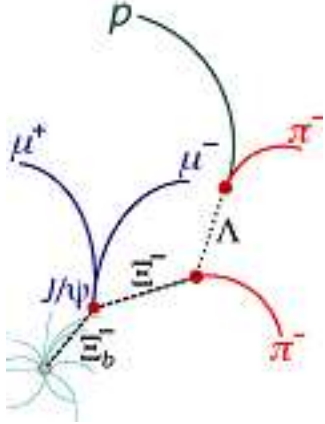}
\caption{The figure shows the decay tree of new particle $\Xi_b^-$,
see ~\cite{:2007ub}.}
\end{figure}
The virtual $W^-$ so
produced then decays into an s- quark and a c- type anti-quark.  The
resulting $c\overline{c}$ then forms a bound state known as the
$J/\Psi$, and leaves a distinctive signature through its decay into
a $\mu^+\mu^-$ pair, while the s- quark produced in the $W^-$ decay 
picks up the spectator d- and s- quarks in the mother $\Xi_b^-$ particle
to form the well-known $\Xi^-$.  This latter then decays when one of
the s- quarks within it decays into $\Lambda$ and $\pi^-$, and then the $\Lambda$ in turn decays into
a proton and a pion.  This sequence of decays lends the name `cascade'
to this class of particles. The total decay can be written as,
$\Xi_b^- \to J/\Psi \Xi^-$ , with $J/\Psi \to \mu^+ \mu^-$ , and $\Xi^- \to \Lambda \pi^- \to p \pi^- \pi^-$.
The D0 collaboration observed 19 candidate events for the production of
these particles.  Based on its own search strategy, the CDF collaboration 
reported 17 candidate events.  The mass determinations are in agreement
with the theoretical predictions based on heavy baryon chiral perturbation
theory, the effective low energy theory of the strong interactions
in this sector, which predicts its mass to be 
$5.806 \pm 0.008 \ {\rm GeV/c^2}$\cite{hep-ph/9609404}.  
Note here that while this particle has spin 1/2 which may be obtained from 
the three constituent quarks each of which carries spin of 1/2, the
theory also predicts a spin 3/2 counterpart that would be denoted by
${\Xi_b^*}^-$, which is yet to be observed.
It may be noted that the D0 results come from $1.3\ {\rm fb^{-1}}$ 
(${\rm fb^{-1}}$ stands for `inverse femtobarn', a standard unit for luminosity)
 integrated
luminostiy, while the CDF results come from $1.9\ {\rm fb^{-1}}$ integrated 
luminosity.
Earlier experiments at  LEP collider at CERN had also reported an indirect evidence 
of $\Xi_b^-$ baryon based on excess of same sign $\Xi^- l^-$ events 
in jets ~\cite{hep-ex/0510023, Phys.Lett.B384.449}.

Not so long ago, according to a Fermilab's Press Release dated October 23, 2006,
the CDF Collaboration discovered the particles denoted by $\Sigma_b^+$
made of b and two u quarks, and also $\Sigma_b^-$ made of b and two
d quarks. Both of these particles carry a spin of 1/2.  The CDF Collaboration
also discovered their spin 3/2 counterparts denoted by
${\Sigma_b^*}^+$ and ${\Sigma_b^*}^-$. CDF results came from integrated luminosity of $1.1  {\rm fb}^{-1}$. These particles have been detected through their sequence of decays; 
$\Sigma_b^{*\pm} \to \Lambda_b^0 \pi^{\pm} $ with $ \Lambda_b^0 \to \Lambda_c^+ \pi^-,  \Lambda_c^+ \to p K^- \pi^+$. 

\newpage

The observed masses are ~\cite{hep-ex/0706.3868} 
\begin{eqnarray*}
m_{\Sigma_b^+} = 5807.8\pm2.0\pm 1.7\ {\rm MeV}
/c^2, & m_{\Sigma_b^{*+}} = 5829.0\pm1.8\pm 1.8\ {\rm MeV}/c^2, \\
m_{\Sigma_b^-} = 5815.2\pm1.0\pm 1.7 \ {\rm MeV}/c^2, & m_{\Sigma_b^{*-}} = 
5836.4\pm2.0\pm 1.8 \ {\rm MeV}/c^2.
\end{eqnarray*}
As expected, the spin 3/2 counterparts
are more massive due to the contributions from the magnetic energy 
costs for aligning the spins of the three quarks to produce this
state, in comparison with the energy costs for the spin 1/2 states. The corresponding predictions from various theories are in agreement with the experimentally observed
masses. 

In summary, noting that there has been significant progress 
in understanding b baryons theoretically, 
up until the discoveries reviewed here, none but
$\Lambda^0_b$ had been observed experimentally. This
particle was already observed in
the early 1980's, for a discussion see ref.~\cite{Griffiths}.
These heavy baryons provide a testing ground for various
approaches to the non-perturbative regime of QCD.  Since
$\Xi^-_b$ has all distinct quarks, it could help in
understanding how individual quarks are bound together inside
hadrons.
With the discoveries reviewed here, a new era of precision
measurements of their properties is now on the horizon.

\end{document}